\DocumentMetadata{
	lang		= eng-US, 	
	pdfstandard	= A-3u,		
	pdfversion	= 1.7, 		
}

\documentclass[colorlinks,upint,nofoot,nohead]{asmeconf} 

\usepackage{enumitem}
\usepackage{booktabs}
\usepackage{tabularx}
\usepackage{siunitx}
\allowdisplaybreaks 

\begin{document}

\ConfName{ }
\ConfAcronym{ }
\ConfDate{ } 
\ConfCity{ }
\PaperNo{ }

%

\title{AI-Driven Performance-to-Design Generation and Optimization of Marine Propellers} 
 
%
%
%

\SetAuthors{%
    Leah Chen\affil{1}, 
	Keni Chih-Hua Wu\affil{2},  
    Boon Tat Chia\affil{1},  
	Xiuqing Xing\affil{2},
    Jian Cheng Wong\affil{2}\CorrespondingAuthor{wongj@a-star.edu.sg}
	}

\SetAffiliation{1}{Mencast Marine, Singapore}
\SetAffiliation{2}{Institute of High Performance Computing, Agency for Science, Technology and Research, Singapore}


\maketitle



\keywords{marine propeller design, generative AI, conditional variational autoencoder, and latent diffusion model, surrogate modeling, evolutionary optimization, data generation.}


\begin{abstract}
Artificial intelligence (AI) is increasingly used to accelerate engineering design by improving decision-making and shortening iteration cycles. Application to marine propeller design, however, remains challenging due to scarce training data and the lack of widely available pretrained models. We address this gap with a physics-based data generation pipeline and a generative-AI framework for direct performance-to-design generation tailored to marine propellers. First, we build a database of over 20{,}000 four- and five-bladed propeller geometries, each accompanied by simulated open-water performance curves, i.e., thrust coefficient, torque coefficient, and open-water efficiency across advance ratios. On top of this dataset, we develop a three-module design framework: (1) A \emph{Conditional Generation Model} that proposes candidate geometries conditioned on design specifications such as target thrust, power, diameter, and blade count. We study both conditional variational autoencoders (cVAE) and latent diffusion models for this task. (2) A \emph{Performance Prediction Model}, implemented as a neural-network surrogate, that predicts thrust, torque, and efficiency in milliseconds, enabling rapid evaluation of generated designs. (3) A \emph{design refinement stage} that applies evolutionary optimization, using either the surrogate or a fast hydrodynamic solver as objective evaluator, to enforce practical constraints such as required thrust under power limits and bounds on blade-area ratio and thickness. Experimental results over a range of operating conditions show that the framework can generate hydrodynamically plausible propeller designs that match prescribed performance targets while substantially reducing design-iteration time relative to the traditional “design spiral” of geometry edits, simulation, and expert-guided refinement. In particular, the latent diffusion–based generator produces more diverse designs under the same conditions, whereas the cVAE tends to yield more uniform solutions, suggesting a stronger capacity for design-space exploration with diffusion models. By coupling physics-based data synthesis with modular AI models, the proposed approach streamlines the propeller design cycle and reduces reliance on expensive high-fidelity simulations to final validation stages.
\end{abstract}







\section{Introduction}

Traditionally, marine propeller design has relied on the expertise of experienced designers, with minor adjustments to standard templates to satisfy specific design needs. This process, described by Praefke~\cite{praefke2011marine} as the “marine propeller design spiral,” involves iterative modeling, simulation, and testing that may take months. High-fidelity simulations, such as those run with Computational Fluid Dynamics (CFD) tool like OpenFOAM, can require days for a single evaluation, making extensive design exploration computationally expensive. Furthermore, small geometric modifications can lead to substantial differences in hydrodynamic performance, further complicating rapid prototyping. Artificial intelligence (AI)-driven surrogate modeling and generative design provide a path to accelerate this process, enabling real-time performance prediction and condition-driven design generation while reducing reliance on repeated high-fidelity simulations.

Inverse design methods based on deep generative models have been extensively explored for two-dimensional (2d) airfoil design, where relatively low-dimensional geometries and abundant benchmark datasets make data-driven approaches attractive. Conditional variational autoencoders (cVAE), generative adversarial networks, and more recently diffusion-based models have been used to correlate the target lift-drag characteristics or pressure distributions directly with the airfoil shapes, learning performance-conditioned geometry manifolds and greatly reducing the need for repeated CFD calls~\cite{yonekura2021data, yonekura2024preliminary, jing2022inverse}. These studies typically operate on 2d airfoils or quasi-3d sections under steady flow, with compact parameterizations and well-curated datasets, forming a mature body of work in aerodynamic inverse design.

In the marine domain, however, most AI and machine-learning efforts have focused on forward surrogate modeling and surrogate-assisted optimization rather than performance-to-design inverse generation. For marine propellers, surrogate models based on Kriging and neural networks have been constructed to predict open-water characteristics from geometric and operating parameters and then coupled with evolutionary or multi-objective optimizers to reduce the computational cost of design studies targeting improved efficiency, mitigated cavitation, and reduced noise~\cite{vesting2014surrogate, jiang2020evolutionary}. Related work on cross-series propeller design and parametric propeller modeling uses machine learning and compact geometric parameterizations to support rapid performance prediction and systematic exploration of the design space, but continues to follow the traditional “geometry $\rightarrow$ performance $\rightarrow$ optimization” loop, without learning a direct performance-to-geometry mapping or leveraging generative models~\cite{tadros2024unified, zheng2023improved}. As highlighted in recent reviews of machine learning for naval architecture and ocean engineering, the dominant use of AI and machine-learning in this field remains prediction and decision support, rather than end-to-end generative design~\cite{panda2023machine}.

In this work, we move from such forward models to a direct performance-to-design paradigm for marine propellers. A key obstacle in applying modern generative and surrogate models to propeller design is the lack of large, standardized datasets and pretrained models, in stark contrast to the situation for airfoils. To overcome this, we first construct a physics-based data generation pipeline that produces a database of more than 20,000 four- and five-bladed propeller geometries, each accompanied by simulated open-water performance curves (thrust coefficient, torque coefficient, and efficiency) over advance ratio. Building on this dataset, we introduce a generative design framework in which (1) conditional generative models (cVAE and latent diffusion) map desired performance specifications to candidate geometries, (2) a fast surrogate model provides real-time estimates of propeller performance, and (3) an evolutionary refinement stage enforces practical design constraints such as target thrust, power limits, and blade-area ratio bounds. Our experimental study demonstrates that the proposed framework can efficiently identify viable designs across different operating conditions while substantially reducing the number and cost of high-fidelity simulations. 

The main contributions of this work are:
\begin{itemize}[noitemsep, topsep=0pt]
    \item \textbf{Physics-based propeller dataset:} We build a dataset of over 20{,}000 four- and five-bladed propellers with full open-water performance curves, providing a reusable asset for data-driven propeller design.
    \item \textbf{Generative performance-to-design mapping:} We develop conditional cVAE and latent diffusion models that generate propeller geometries directly from specified performance targets, replacing the traditional “geometry $\rightarrow$ performance $\rightarrow$ optimization” loop. 
    \item \textbf{Fast, high-accuracy surrogate model:} We train a performance-prediction surrogate that evaluates propeller thrust, torque, and efficiency in milliseconds, enabling interactive analysis and large-scale design exploration.
    \item \textbf{Constraint-aware design refinement:} We couple the generative models with an evolutionary refinement stage that enforces practical constraints such as target thrust, power limits, and blade-area ratio bounds using the surrogate (or a fast solver) as objective evaluator.
    \item \textbf{Empirical evaluation and model comparison:} Our experimental results show that the framework quickly finds viable designs under varied operating conditions and that the diffusion-based generator yields more diverse yet performant designs than the cVAE.
\end{itemize}

\section{Problem Formulation}

\begin{figure}[t]
\centering
\includegraphics[width=0.45\columnwidth]{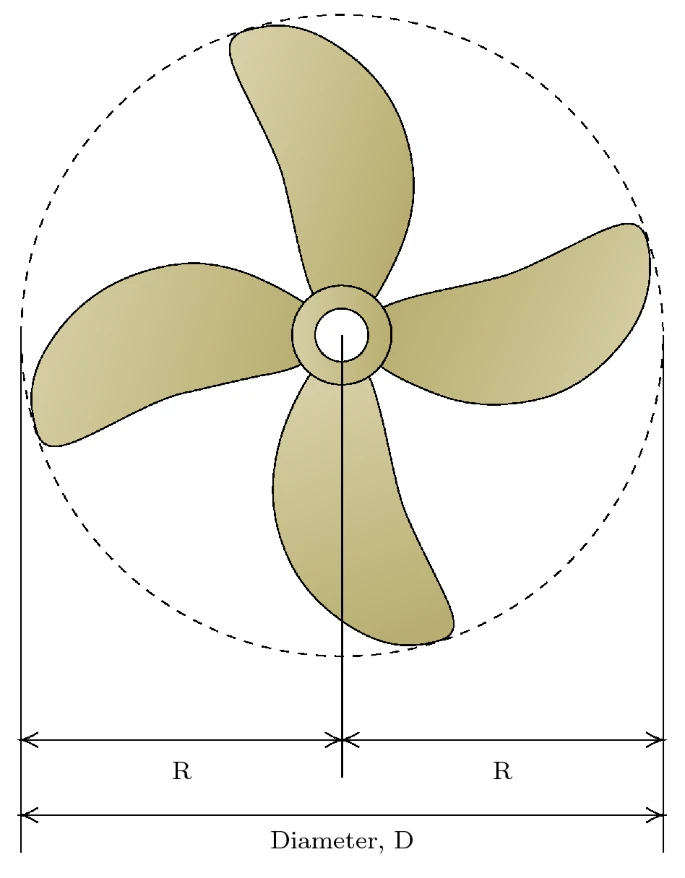}
\hfill
\includegraphics[width=0.5\columnwidth]{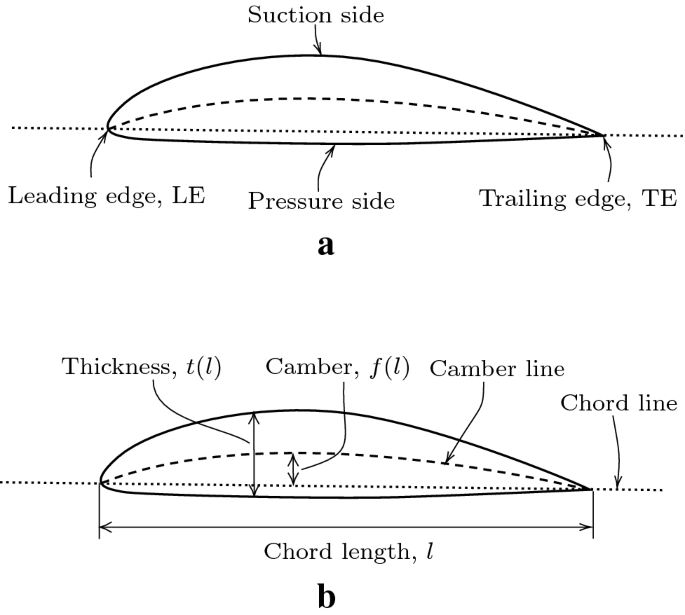}
\caption{(Left) diameter, (right) chord length, maximum thickness, and maximum camber~\cite{njaastad2022identification}.}
\label{method:propeller:geometry}
\end{figure}

\begin{figure*}[!ht]
\centering
\includegraphics[width=0.75\columnwidth]{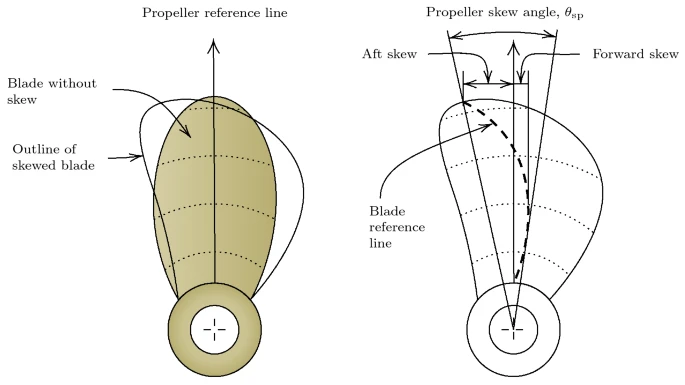}
\hfill
\includegraphics[width=0.65\columnwidth]{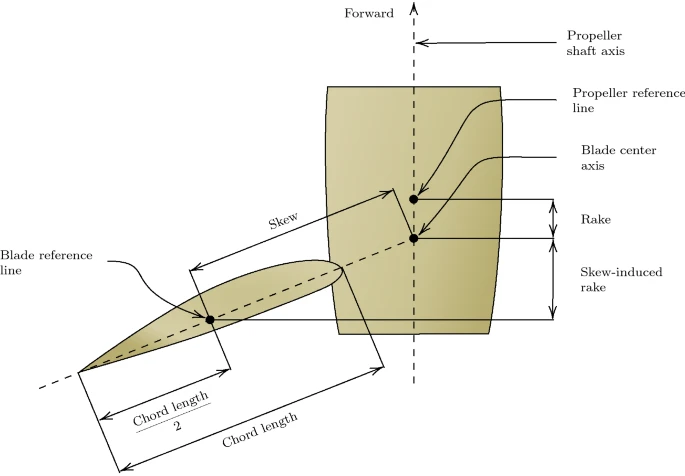}
\hfill
\includegraphics[width=0.55\columnwidth]{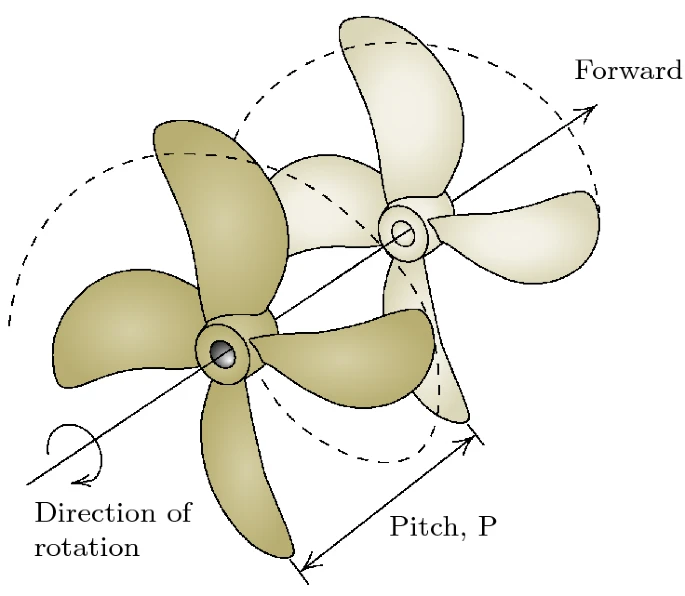}
\caption{(Left) skew, (middle) rake, and (right) pitch~\cite{njaastad2022identification}.}
\label{method:propeller:skew_rake_pitch}
\end{figure*}

\subsection{Propeller Geometry Representation} \label{sec:geometry_parameterization}

We parameterize each 3d propeller geometry by six radial distributions of localized geometric features, i.e., chord length, skew, maximum thickness, rake, pitch, and maximum camber, sampled at 27 normalized radial positions, yielding a 162-dimensional feature vector. \textit{Chord length} (Figure~\ref{method:propeller:geometry}) describes the width of the blade section at a given radius. For a fixed diameter and operating condition, larger chord increases the available pressure surface, which generally raises the thrust-carrying capacity and cavitation margin. Conversely, smaller chord increases loading per unit area, which can improve efficiency in lightly loaded regions but may lead to higher peak pressures and cavitation if pushed too far. The chord distribution along the radius therefore plays a central role in shaping the thrust distribution, expanded area ratio, and overall open-water characteristics. \textit{Skew} (Figure~\ref{method:propeller:skew_rake_pitch}) is the angular displacement of the blade’s reference line relative to the hub center. Skew is important for influencing cavitation cavitation behaviour and propulsive efficiency~\cite{carlton2018marine}. \textit{Maximum thickness} and \textit{maximum camber} (Figure~\ref{method:propeller:geometry}) describe the foil shape of each blade section and directly affect the local lift and drag characteristics. \textit{Rake} (Figure~\ref{method:propeller:skew_rake_pitch}) is defined as the axial displacement of a blade section along the shaft line. Positive rake can reduce propeller–hull interaction and modify the thrust distribution. \textit{Pitch} (Figure~\ref{method:propeller:skew_rake_pitch}) is the theoretical distance a propeller would advance in one full revolution in a solid medium and strongly influences both thrust generation and efficiency.

In addition to these radial features, we include global parameters: propeller diameter $D$ (Figure~\ref{method:propeller:geometry}) and blade number $B$. Together, these descriptors form a compact yet expressive representation that is standard in propeller hydrodynamics and can be recovered from 3d scans of existing propellers~\cite{njaastad2022identification}. The parameterization is also fully compatible with the input format of the hydrodynamic simulation software~\textsc{PropElements}\footnote{\textsc{PropElements} is developed by HydroComp, Inc.}, which facilitates validating both simulated and AI-generated designs within a consistent computational framework.

\subsection{Performance and Constraints}

\begin{figure}[t]
\centering
\includegraphics[width=1.\columnwidth]{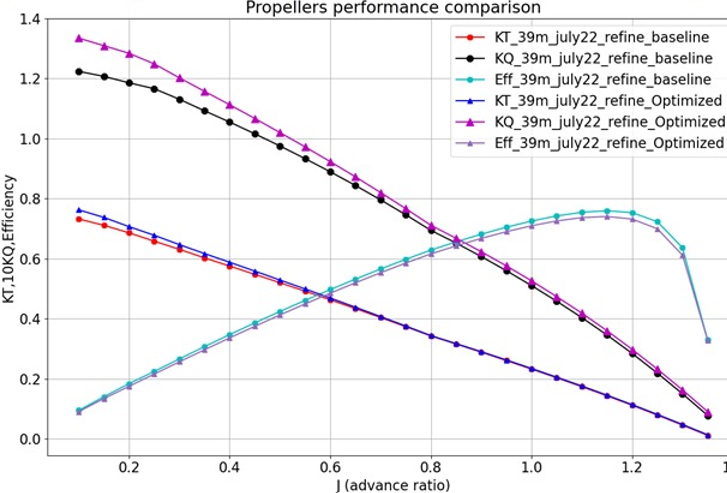}
\caption{Open-water characteristics of two propellers.}
\label{method:fig:propeller_performance}
\end{figure}

The hydrodynamic performance of a propeller is conventionally described by its open-water characteristics, i.e., thrust coefficient $K_T$, torque coefficient $K_Q$, and open-water efficiency $\eta$ as functions of the advance ratio $J$ (Figure~\ref{method:fig:propeller_performance}). The \emph{advance ratio} $J$ is defined as $J = \frac{V_A}{nD}$, where $V_A$ is the advance speed (m/s), $n$ is the shaft rotation speed (revolutions per second), and $D$ is the propeller diameter (m). For a given $J$, the corresponding non-dimensional performance coefficients are $K_T = \frac{T}{\rho n^2 D^4}$, $K_Q = \frac{Q}{\rho n^2 D^5}$, and $\eta = \frac{J}{2\pi} \,\frac{K_T}{K_Q}$, where $T$ is thrust, $Q$ is torque, and $\rho$ is the water density. The open-water efficiency $\eta$ measures how effectively the propeller converts torque into thrust when operating in a uniform inflow, without hull interaction effects.

The open-water performance curves $K_T(J)$, $K_Q(J)$, and $\eta(J)$ form the primary design targets and constraints for propeller designers. In practice, operating conditions such as ship speed and required thrust are specified together with limits on propeller diameter, available engine power, and shaft revolutions per minute (RPM). The design task is then to identify a propeller whose open-water characteristics satisfy these requirements at the intended operating point. 

\subsection{Inverse Performance-to-Design Problem}

In our framework, performance targets and constraints directly define the conditioning inputs used in the inverse design and optimization procedure. We cast this as a performance-to-design inverse problem. At a given service condition, the ship (or propeller) designer typically specifies:
\begin{itemize}[noitemsep, topsep=0pt]
    \item advance speed $V_A$ (m/s), i.e., ship speed at the propeller plane,
    \item required thrust $T_{\text{req}}$ (N) to balance hull resistance,
    \item shaft rotation speed $n$ (rev/s), i.e., shaft RPM$\div$60,
    \item available shaft (engine) power $P_{\text{avail}}$ (W),
    \item allowable propeller diameter $D$ (m), and discrete design choices such as blade number $B$.
\end{itemize}
These quantities determine a \emph{target operating point} on the open-water curves. First, the corresponding advance ratio is
\begin{equation}
    J^\star = \frac{V_A}{n D}.
\end{equation}
Given the required thrust, the corresponding target thrust coefficient at this operating point is
\begin{equation}
    K_T^\star(J^\star) = \frac{T_{\text{req}}}{\rho n^2 D^4},
\end{equation}
where $\rho$ is the water density. The available shaft power $P_{\text{avail}}$ and shaft speed $n$ define the maximum admissible torque, $Q_{\text{max}} = \frac{P_{\text{avail}}}{2\pi n}$, and thus an associated torque coefficient
\begin{equation}
    K_Q^\star(J^\star) = \frac{Q_{\text{max}}}{\rho n^2 D^5}
    = \frac{P_{\text{avail}}}{2\pi \rho n^3 D^5}.
\end{equation}
Combining these, the implied open-water efficiency at the design point is
\begin{equation}
    \label{eq:open_water_eta}
    \eta^\star(J^\star) = \frac{J^\star}{2\pi} \,\frac{K_T^\star}{K_Q^\star},
\end{equation}
which expresses how effectively the propeller can, in principle, convert the available torque into thrust at the specified condition.

The region of \emph{optimal performance} is typically associated with the range of advance ratios where the propeller can deliver the required thrust while maintaining high open-water efficiency. In marine engineering practice, designs are tuned so that the service operating point lies near this peak-efficiency region, subject to constraints on delivered thrust, power absorption, cavitation margin, and geometric limitations such as maximum diameter, blade-area ratio, and thickness. In the present study, however, we focus on a single primary operating point for each design and do not explicitly account for off-design performance at this stage.

The inverse design problem we address can therefore be stated as: given a set of target quantities $(V_A, T_{\text{req}}, n, P_{\text{avail}}, D)$ defining $(J^\star, K_T^\star, K_Q^\star, \eta^\star)$, find the 3d propeller geometry whose open-water characteristics match these targets as closely as possible while satisfying the imposed conditions such as the diameter $D$ and blade number $B$.

\section{Physics-Based Data Generation}

To enable AI model training, we first construct a synthetic propeller database of more than 20{,}000 four- and five-bladed geometries, each accompanied by simulated open-water performance curves $K_T(J)$, $K_Q(J)$, and $\eta(J)$ over appropriate range of advance ratio $J$. This dataset provides the paired geometry-performance information required for both generative performance-to-design mapping and fast surrogate modeling.

We sample propeller geometries in a compact latent space derived from expert-designed propellers. As described in Section~\ref{sec:geometry_parameterization}, each geometry is represented by six radial distributions of localized geometric features (chord, skew, maximum thickness, rake, pitch, and maximum camber) sampled at 27 normalized radial positions. We first collect a set of expert or reference propeller designs (with further details summarized Appendix~\ref{app:reference_designs} and  Table~\ref{tab:hull_summary}) and apply principal component analysis (PCA) to these 162-dimensional vectors, yielding a lower-dimensional latent embedding that captures the dominant modes of geometric variation. New synthetic geometries are then generated by randomly sampling in this latent space and mapping the samples back to the original 162-dimensional space via the inverse PCA transform. In parallel, we randomly assign the blade number $b \in \{4,5\}$ and sample the propeller diameter $D \in [0.5,2.5]$. For each sampled geometry, we evaluate open-water performance using \textsc{PropElements}. 

We then organize the data into two task-specific formats:
\begin{itemize}[noitemsep, topsep=0pt]
    \item \textbf{Generative performance-to-design mapping.} For training the conditional generative models (cVAE and latent diffusion), each sample consists of the 162-dimensional design vector and a conditioning vector that summarizes the operating condition and performance, e.g., $\mathbf{c}_{\text{gen}} = [J, K_T, K_Q, \eta_, D, B]$. Conditioning on $\mathbf{c}_{\text{gen}}$ allows the generative model to propose geometries that meet specified performance and operating requirements. Dataset splits were 80\% training and 20\% validation for generative models.
    \item \textbf{Performance prediction surrogate model.} For training the surrogate model, each sample includes the 162-dimensional design vector augmented with the global parameters $D$, $B$, and the operating advance ratio $J$, yielding a 165-dimensional input, together with the corresponding target performance metrics $[K_T, K_Q, \eta].$ This formulation enables the surrogate to predict performance at arbitrary operating points for a given geometry. The dataset comprises 278{,}760 samples, which are randomly split into 70\% training, 15\% validation, and 15\% test subsets.
\end{itemize}
Data were standardized prior to AI model training.

\section{AI-Driven Performance-to-Design Framework}

\subsection{Conditional Generation Models}

The inverse performance-to-design problem is inherently one-to-many: many different geometries can exhibit very similar open-water performance at a given operating point. A purely deterministic machine learning model would tend to collapse this variability into a single ``average'' solution and cannot represent alternative, yet equally valid, designs. Instead, we adopt \emph{conditional generative models} that learn the full conditional distribution $p(x \mid \mathbf{c}_{\text{gen}})$ of geometries $x$ given performance and operating conditions $\mathbf{c}_{\text{gen}}$. This allows us to draw multiple candidate designs that all meet the same performance targets, while still giving the designer explicit control over the conditioning variables.

The open-water relations couple the three performance coefficients via Eq.~\eqref{eq:open_water_eta}. For a fixed $J$, the pair $(K_T,\eta)$ uniquely determines $K_Q$. Including $K_Q$ explicitly in the conditioning vector would therefore introduce redundancy and potential multicollinearity. We thus condition on the compact vector $\mathbf{c}_{\text{gen}} = [J, K_T, \eta, D, B]$ and recover $K_Q$ as needed, preserving a full performance specification while keeping the conditioning low-dimensional.

\subsubsection{Conditional Variational Autoencoder (cVAE).}

The cVAE operates on the 162-dimensional design vector $x \in \mathbb{R}^{162}$ described in Section~\ref{sec:geometry_parameterization}, and uses the conditioning vector $\mathbf{c}_{\text{gen}}$ to steer generation towards specified performance targets and operating conditions. The encoder $q_\phi(z \mid x, \mathbf{c}_{\text{gen}})$, i.e., a neural network parameterized by $\phi$, maps the input design and condition to a latent distribution over $z \in \mathbb{R}^{d_z}$, while the decoder $p_\theta(x \mid z, \mathbf{c}_{\text{gen}})$, i.e., a neural network parameterized by $\theta$, reconstructs a design conditioned on both the latent variable and the same condition vector. The prior for the latent variable is assumed to be standard normal, i.e., \ $p(z \mid \mathbf{c}_{\text{gen}}) = \mathcal{N}(0, I)$, independent of the condition.

The cVAE is trained by minimizing a $\beta$-weighted conditional evidence lower bound (ELBO),
\begin{equation}
    L_{\text{cVAE}} =
    \mathbb{E}_{q_\phi(z \mid x, \mathbf{c}_{\text{gen}})}
    \big[ \| x - \hat{x} \|_2^2 \big]
    + \beta \, D_{\mathrm{KL}}\big( q_\phi(z \mid x, \mathbf{c}_{\text{gen}}) \,\big\|\, p(z \mid \mathbf{c}_{\text{gen}}) \big),
    \label{eq:cvae_loss}
\end{equation}
where $\hat{x}$ is the mean of $p_\theta(x \mid z, \mathbf{c}_{\text{gen}})$, i.e., the reconstructed design, and $\beta$ controls the trade-off between reconstruction fidelity and latent regularization. Here, $D_{\mathrm{KL}}(q \,\|\, p)$ denotes the Kullback–Leibler divergence between two distributions $q$ and $p$, measuring how far the encoder distribution $q_\phi(z \mid x, \mathbf{c}_{\text{gen}})$ deviates from the prior $p(z \mid \mathbf{c}_{\text{gen}})$.

At inference time, we sample $z \sim \mathcal{N}(0,I)$ and decode with a chosen condition $\mathbf{c}_{\text{gen}}$ via $p_\theta(x \mid z, \mathbf{c}_{\text{gen}})$, which yields diverse geometries consistent with the same performance target.

\subsubsection{Latent Diffusion Model.}

We also explore diffusion-based model for the performance-to-design generation task. As a baseline, we first trained a conditional diffusion model directly on the 162-dimensional design vectors using a 1D residual backbone. While this model was able to capture global statistics of the training set, the generated designs exhibited irregular radial trends (e.g., non-smooth chord and pitch distributions) and inconsistencies across propeller sizes, highlighting the limitations of applying diffusion directly to high-dimensional flattened feature vectors.

To improve geometric coherence, we adopt a latent diffusion approach~\cite{rombach2022high}. A variational autoencoder (VAE) is first trained to compress the 162-dimensional design vector $x$ into a lower-dimensional latent representation $z$:
\[
    x \in \mathbb{R}^{162}
    \xrightarrow{\ \text{Encoder}\ }
    z \in \mathbb{R}^{{d_z}}.
\]
The VAE is optimized with a standard reconstruction-plus-KL objective,
\begin{equation}
    L_{\text{VAE}} =
    \mathbb{E}_{q_\psi(z \mid x)}
    \big[ \| x - \hat{x} \|_2^2 \big]
    + \beta_{\text{VAE}} \, D_{\mathrm{KL}}\big( q_\psi(z \mid x) \,\big\|\, \mathcal{N}(0,I) \big),
\end{equation}
where $\hat{x}$ is the decoded reconstruction and $\beta_{\text{VAE}}$ is a weighting factor. Once trained, the VAE parameters are frozen and only the encoder–decoder pair is used to move between design space and latent space.

A conditional diffusion model is then trained in the latent space. At each diffusion timestep $t$, the model takes as input a noisy latent code $z_t$ and the conditioning vector $\mathbf{c}_{\text{gen}} = [J, K_T, \eta, D, B]$, and predicts either the added noise or a velocity vector in latent space. The training objective can be written in the noise-prediction form as
\begin{equation}
    L_{\text{diffusion}} =
    \mathbb{E}_{z_0, t, \epsilon}
    \big[ \| \epsilon - \epsilon_\theta(z_t, t, \mathbf{c}_{\text{gen}}) \|_2^2 \big],
\end{equation}
where $z_0$ is the encoded latent of a clean design, $\epsilon \sim \mathcal{N}(0,I)$ is Gaussian noise, and $\epsilon_\theta$ is the denoising network. In practice, we experiment with both noise-prediction and velocity-prediction variants; both follow the same conditional training protocol in latent space.

At inference time, we start from a Gaussian noise in latent space, iteratively denoise conditioned on $\mathbf{c}_{\text{gen}}$, and decode the final latent sample back to a 162-dimensional propeller geometry via the VAE decoder.

\subsection{Performance Prediction Surrogate Model}

The forward prediction task aims to estimate hydrodynamic performance metrics $(K_T, K_Q, \eta)$ given a propeller design and operating condition. Each sample for the surrogate model is represented by a 165-dimensional input vector $\mathbf{x}_{\text{surr}} = [\text{design}_{162}, D, B, J]$ where $\text{design}_{162}$ is the 162-dimensional design vectors, $D$ is the propeller diameter, $B$ is the blade count, and $J$ is the advance ratio. The corresponding target output is $mathbf{y}_{\text{surr}} = [K_T, K_Q, \eta]$

We model the mapping $\mathbf{x}_{\text{surr}} \mapsto \mathbf{y}_{\text{surr}}$ with a fully connected multilayer perceptron (MLP) with dropouts to improve robustness and reduce overfitting. The MLP is trained using the mean squared error (MSE) loss and the Adam optimizer:
\begin{equation}
    L_{\text{MSE}} =
    \frac{1}{N} \sum_{i=1}^N \big\| \mathbf{y}_i - \hat{\mathbf{y}}_i \big\|_2^2,
\end{equation}
where $\mathbf{y}_i$ denotes the simulated performance vector for sample $i$, $\hat{\mathbf{y}}_i$ the network prediction, and $N$ the number of training samples. For evaluation on the test set, we report the root-mean-squared error (RMSE), the relative $L_2$ error, and the coefficient of determination $R^2$.

\subsection{Design Refinement with Evolutionary Optimization}

While the conditional generative models provide diverse candidate geometries that broadly satisfy the specified performance targets, practical propeller design must also account for additional engineering constraints and preferences. These include blade-area ratio (BAR) and thickness, as well as considerations related to cavitation risk. In practice, BAR bounds are typically set based on prior experience and client preference (e.g., tug boats often target BAR$=0.6-0.8$, while faster craft may use BAR$=0.8-1.2$), whereas blade thickness is treated as a hard constraint following classification-rule minimum thickness requirements (see Appendix~\ref{app:thickness_constraint}). To incorporate such requirements, we introduce a \emph{design refinement} stage based on evolutionary optimization.

In this stage, the designs proposed by the generative models are used as starting points for a continuous optimization process in the design space. We employ a covariance-matrix adaptation evolution strategy (CMA-ES) algorithm~\cite{hansen2006cmaes}, which is well suited to high-dimensional, non-convex problems and does not rely on gradient information. The optimizer iteratively perturbs candidate geometries, evaluates their hydrodynamic performance, and updates its search distribution toward designs that better satisfy the targets and constraints.

The refinement loop is designed to flexibly use either the trained performance-prediction model or a fast hydrodynamic solver as the objective evaluator. The surrogate enables rapid exploration and ranking of a large number of candidates, whereas the solver can be employed selectively when higher-fidelity assessment is required. Within this framework, we can enforce practical constraints such as achieving the required thrust under a tighten power budget, maintaining BAR within user-specified bounds, enforcing minimum blade thickness (Appendix~\ref{app:thickness_constraint}), and keeping cavitation-related indicators within acceptable limits, at the cost of a higher-latency search compared to direct sampling. As a result, the evolutionary refinement stage acts as a bridge between the data-driven generative models and engineering practice, turning promising AI-generated concepts into designs that are more immediately usable in a real propeller design workflow.

\section{Experimental Study and Results}

\subsection{Model Training Setup and Validation}

All AI models were trained on an NVIDIA RTX 5090 GPU. Validation of the performance-to-design generative models followed a two-stage procedure. First, we performed evaluation using the performance prediction surrogate model: for each conditioning vector, multiple designs were generated and evaluated by the MLP to check whether the predicted $(K_T, K_Q, \eta_O)$ at the target advance ratio matched the conditioning specification within a prescribed tolerance. This step enabled efficient screening of a large number of generated samples. Second, selected designs were further validated using \textsc{PropElements} to ensure hydrodynamic plausibility and to quantify any surrogate-induced bias.

\begin{figure}[!htb]
\centering
\includegraphics[width=0.45\columnwidth]{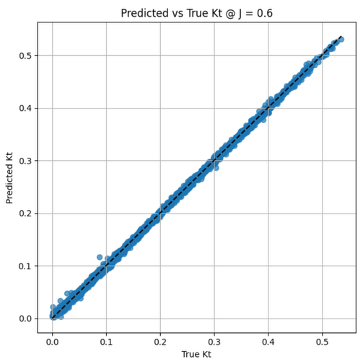}
\hfill
\includegraphics[width=0.45\columnwidth]{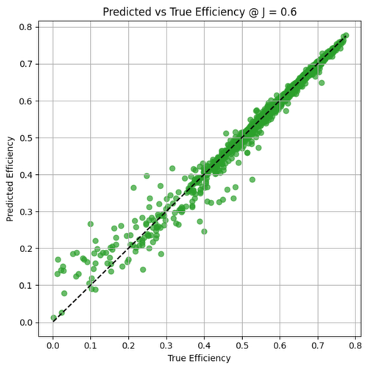}
\caption{Predicted vs. actual for (left) thrust coefficient $K_T$ and (right) efficiency $\eta$, at $J$=0.6.}
\label{result:fig:MLP_scatter}
\end{figure}

\subsection{Performance Prediction Model}

Test-set results for the performance-prediction model are summarized in Table~\ref{tab:forward_pred}. The MLP surrogate achieved high coefficients of determination $R^2$ for all three targets $(K_T, K_Q, \eta)$, together with low RMSE and relative $L_2$ errors, indicating strong predictive capability across the sampled design and operating space. Among the three metrics, open-water efficiency $\eta$ was the most challenging to predict, exhibiting a slightly lower $R^2$ and higher relative error than $K_T$ and $K_Q$. This can be attributed in part to data imbalance: designs with very low efficiency (e.g.\ $\eta < 0.4$) are underrepresented in the dataset (Figure~\ref{result:fig:MLP_scatter}), and errors in these regions have limited impact on practical design, where attention is primarily focused on high-efficiency operating conditions.

\begin{table}[!htb]
\centering
\caption{Accuracy of MLP model on test dataset.}
\label{tab:forward_pred}
\begin{tabular}{lccc}
\hline
Metric & $R^2$ & Relative $L_2$ & RMSE \\
\hline
$K_T$ & 0.9990 & 0.0167 & 0.0051 \\
$K_Q$ & 0.9988 & 0.0207 & 0.0011 \\
$\eta$ & 0.9946 & 0.0314 & 0.0164 \\
\hline
\end{tabular}
\end{table}

\begin{figure}[!htb]
\centering
\includegraphics[width=1.0\columnwidth]{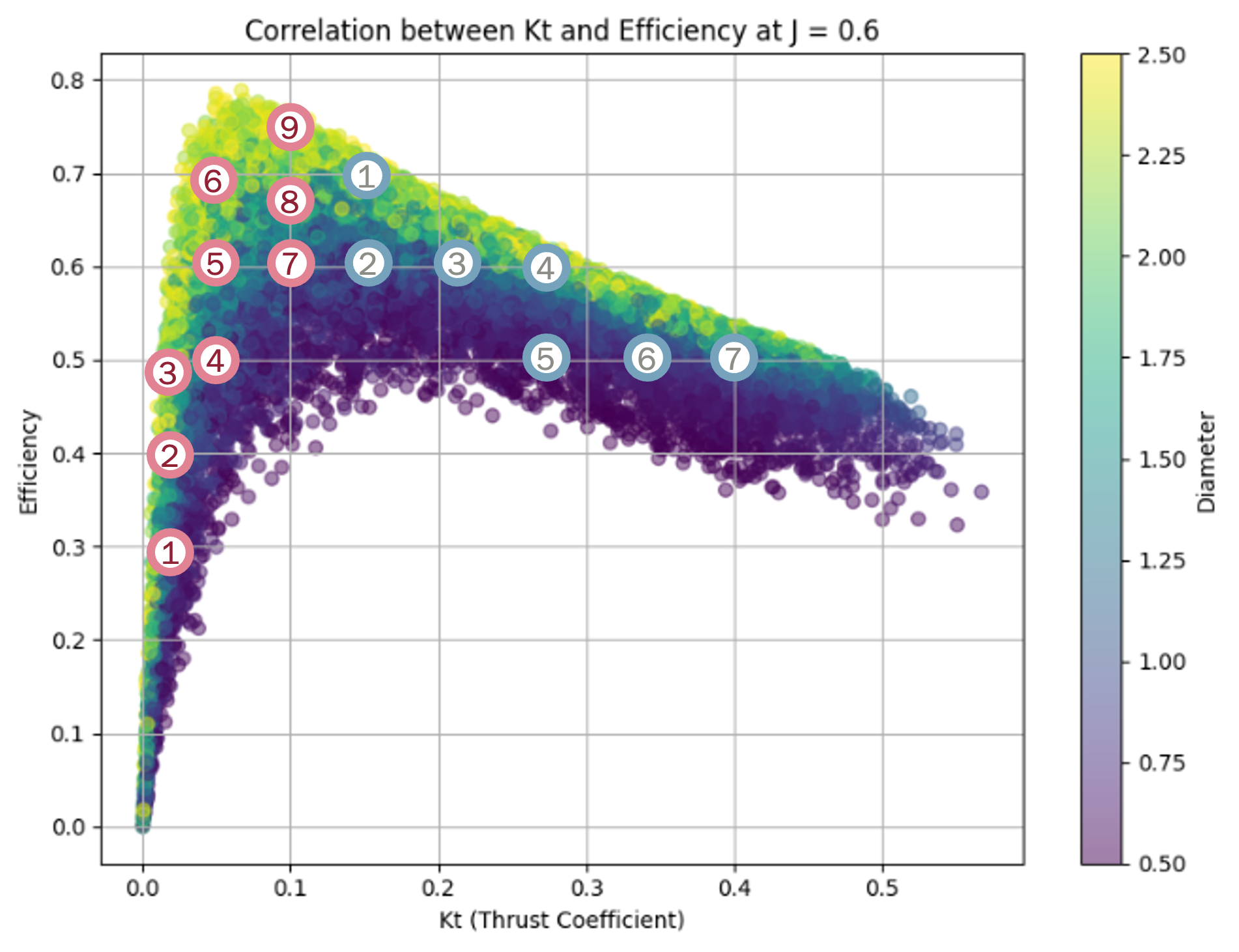}
\caption{Distribution of $K_t$ and $\eta$ at $J=0.6$ with target condition points. Pink labels: efficiency coverage; blue labels: $K_T$ coverage. Color scale: diameter is in meters.}
\label{result:fig:condition_vectors}
\end{figure}

\begin{table}[!htb]
\centering
\caption{Generated-sample errors for cVAE models with different $\beta$ (MLP validation).}
\small
\label{tab:cVAE_beta}
\begin{tabular}{lcccc}
\hline
$\beta$ & $K_T$ Error \% & $\eta$ Error \% & $K_T$ $L_2$ Error & $\eta$ $L_2$ Error \\
\hline
0.50 & 0.8571 & 4.3487 & 0.2521 & 2.4053 \\
0.10 & 0.9880 & 4.9435 & 0.3208 & 2.8428 \\
0.07 & 1.2367 & 6.3648 & 0.3448 & 3.6408 \\
0.05 & 1.2679 & 5.7202 & 0.3667 & 3.6286 \\
0.02 & 1.3194 & 4.1839 & 0.3788 & 2.5053 \\
\hline
\end{tabular}
\end{table}

\begin{figure}[!htb]
\centering
\includegraphics[width=1\columnwidth]{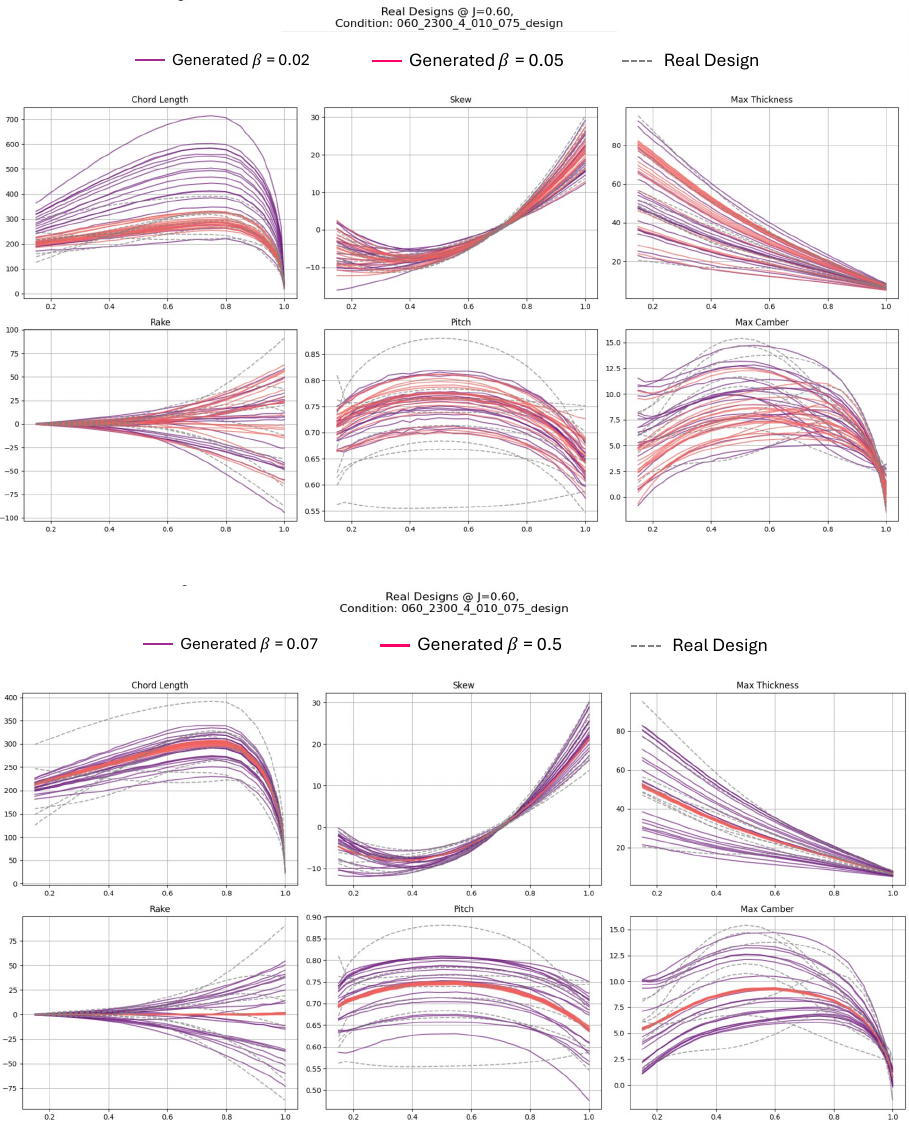}
\vspace{-2mm}
\caption{Generated designs from the cVAE model trained with different $\beta$ values. Upper figure shows small $\beta$ values ($0.02$ and $0.05$), allowing more diverse generated designs. Lower figure shows larger $\beta$ values ($0.07$ and $0.5$), which increases similarity across generated designs.}
\label{result:fig:cvae_beta}
\end{figure}

\subsection{Conditional Variational Autoencoder (cVAE) Model}

A key hyperparameter in cVAE training is the weight $\beta$ in Eq.~\eqref{eq:cvae_loss}, which balances reconstruction accuracy against KL-divergence regularization of the latent space. Smaller values of $\beta$ encourage the model to focus on reconstructing the training designs, while larger values place more emphasis on aligning the encoder distribution with the prior, which can improve latent-space structure but may reduce sample diversity or fidelity.

We systematically studied the effect of $\beta$ by training a family of cVAE models and evaluating the performance of generated designs via the forward MLP surrogate. For this study, we fixed the advance ratio at $J = 0.6$ (the most common operating condition in the dataset) and constructed 16 distinct conditioning vectors (different combinations of $K_T$, $\eta$, and diameter $D$, with $B \in \{4,5\}$; see Figure~\ref{result:fig:condition_vectors}). For each condition, we generated 20 designs, resulting in 640 samples per model configuration. Table~\ref{tab:cVAE_beta} reports the resulting errors in $K_T$ and $\eta$ as estimated by the surrogate.

In general, increasing $\beta$ within a moderate range reduced the $L_2$ errors in both $K_T$ and $\eta$, but very large values yielded low sample diversity: for $\beta > 0.07$, the cVAE produced nearly identical outputs across repeated designs for the same condition (Figure~\ref{result:fig:cvae_beta}). Conversely, very small values ($\beta < 0.02$) occasionally produced designs that fell outside the distribution of geometries seen during training. Based on this trade-off, we identified $\beta \in [0.02, 0.07]$ as a suitable range and subsequently refined the learning rate within this band. After several iterations, we fixed $\beta = 0.07$ and conducted a learning-rate tuning, evaluating each trained cVAE model using both the MLP surrogate and \textsc{PropElements}. Both evaluators indicated that a learning rate of $1 \times 10^{-4}$ provided the best overall performance (Tables~\ref{tab:cVAE_lr_MLP} and~\ref{tab:cVAE_lr_propelement}), leading to our final model.

\begin{figure}[!htb]
\centering
\includegraphics[width=0.49\columnwidth]{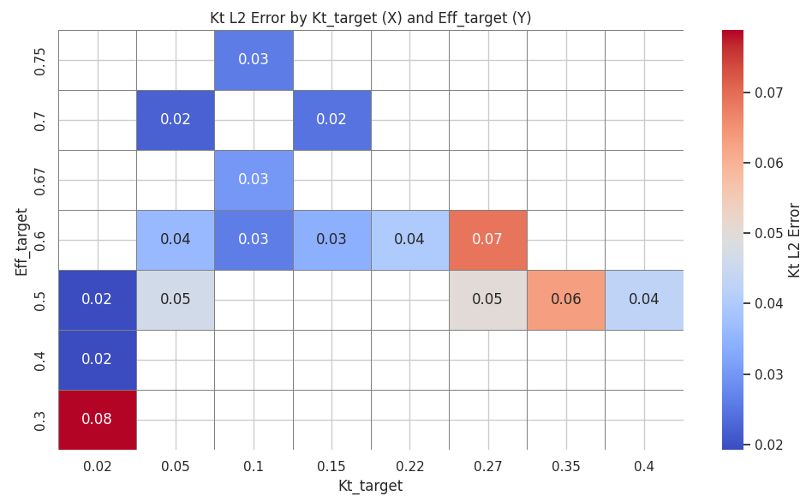}
\hfill
\includegraphics[width=0.49\columnwidth]{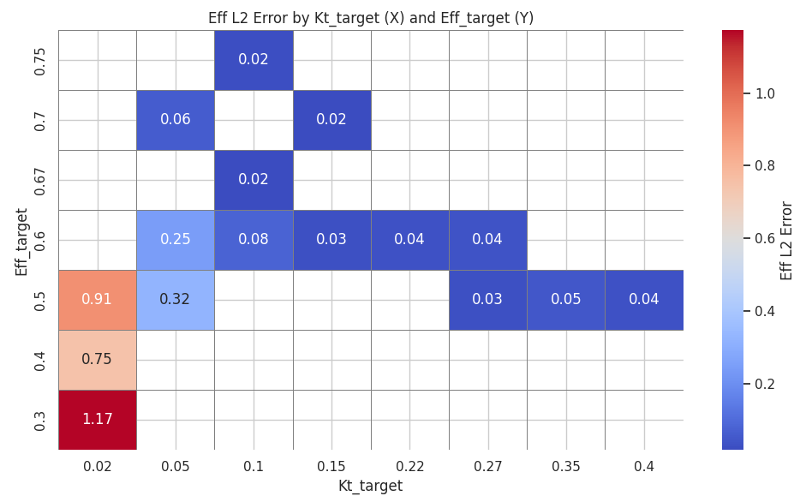}
\caption{(Left) relative $L_2$ error for $K_T$ across target conditions at $J=0.6$. Higher error at $K_T > 0.25$. (Right) relative $L_2$ error for $\eta$ across target conditions at $J=0.6$. Higher error at $\eta < 0.05$.}
\label{result:fig:cVAE_errors}
\end{figure}

\begin{figure}[!htb]
\centering
\includegraphics[width=1.0\columnwidth]{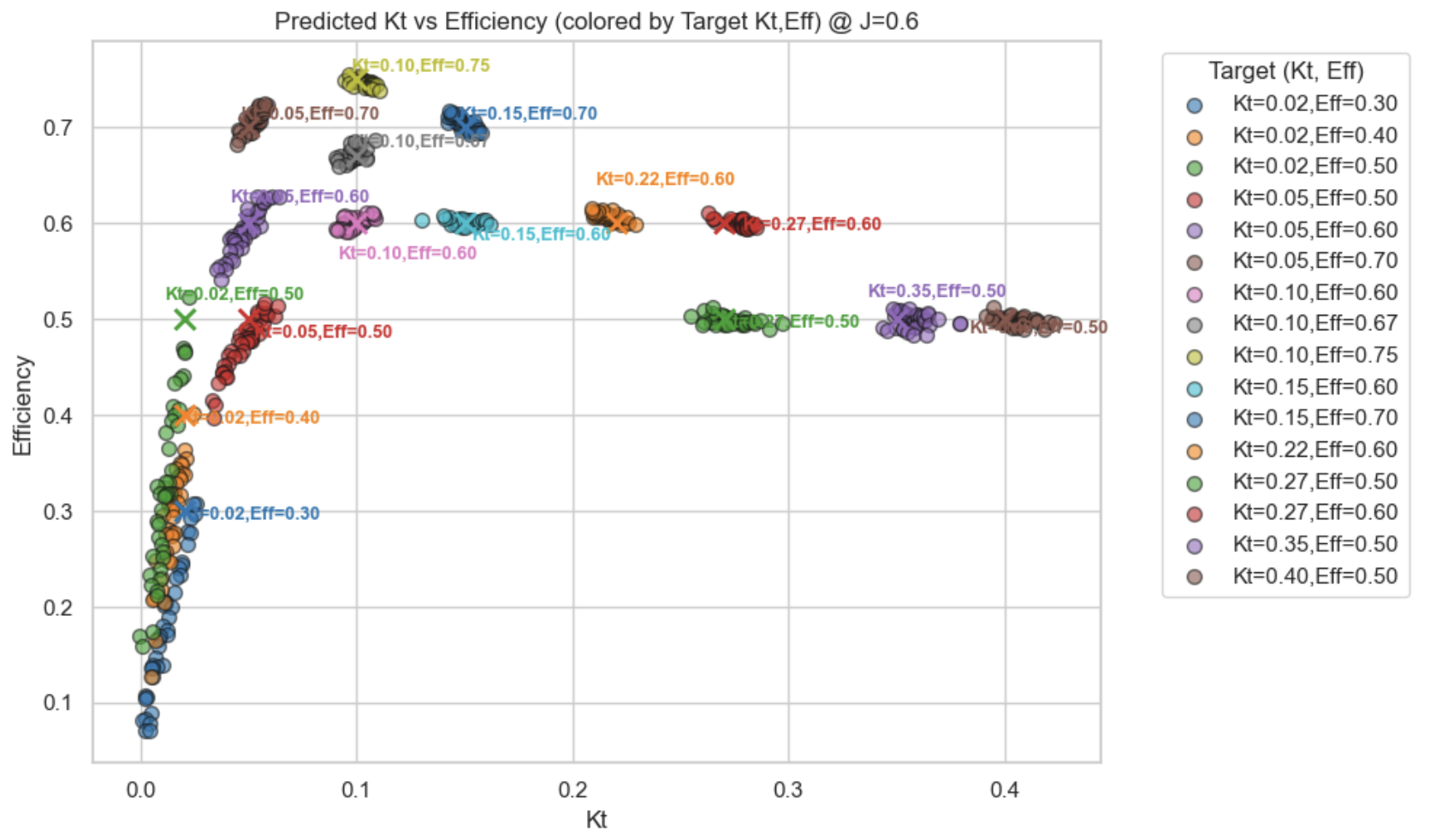}
\caption{Distribution of designs generated by cVAE. The “X” marks indicate the target $(K_T,\eta)$ conditions.}
\label{result:fig:cVAE_validation}
\end{figure}

The model errors in $K_T$ and $\eta$ over the test conditions are shown in Figure~\ref{result:fig:cVAE_errors}. At $J = 0.6$, conditions with high loading ($K_T > 0.25$) and very low efficiency ($\eta < 0.05$) exhibited noticeably larger discrepancies, which we attribute to data sparsity in these extreme regions of the operating space. This pattern is further illustrated in Figure~\ref{result:fig:cVAE_validation}, where deviations from the targets are more pronounced at low $\eta$ (vertical spread) and high $K_T$ (horizontal spread). Overall, the cVAE achieves good alignment with the conditioning targets in the practically relevant, high-efficiency regime, with most of the residual error concentrated in low-efficiency regions that are less critical for typical design tasks.

\begin{table}[t]
\centering
\caption{Errors in matching conditioning targets for cVAE models with different learning rates (evaluated by MLP surrogate).}
\resizebox{\columnwidth}{!}{%
\begin{tabular}{lcccc}
\hline
LR & $K_T$ Error \% & $\eta$ Error \% & $K_T$ $L_2$ Error & $\eta$ $L_2$ Error \\
\hline
$5\times 10^{-4}$ & 1.0285 & 5.9827 & 0.2000 & 2.3688 \\
$1\times 10^{-4}$ & \textbf{0.9629} & \textbf{3.8224} & \textbf{0.1814} & \textbf{1.5740} \\
$5\times 10^{-5}$ & 1.2982 & 4.3213 & 0.2723 & 1.7540 \\
$1\times 10^{-5}$ & 1.4361 & 5.2170 & 0.2760 & 1.7580 \\
\hline
\end{tabular}%
}
\label{tab:cVAE_lr_MLP}
\end{table}

\begin{table}[t]
\centering
\caption{Errors in matching conditioning targets for cVAE models with different learning rates (evaluated by \textsc{PropElements}).}
\resizebox{\columnwidth}{!}{%
\begin{tabular}{lcccc}
\hline
LR & $K_T$ Error \% & $\eta$ Error \% & $K_T$ $L_2$ Error & $\eta$ $L_2$ Error \\
\hline
$5\times 10^{-4}$ & 0.9475 & 2.9863 & 0.1816 & 1.2082 \\
$1\times 10^{-4}$ & \textbf{0.7594} & \textbf{1.5710} & \textbf{0.1536} & \textbf{0.7329} \\
$5\times 10^{-5}$ & 1.1985 & 1.7292 & 0.2593 & 0.7134 \\
$1\times 10^{-5}$ & 1.3066 & 4.3070 & 0.2528 & 1.2968 \\
\hline
\end{tabular}%
}
\label{tab:cVAE_lr_propelement}
\end{table}

\begin{figure}[!hb]
\centering
\includegraphics[width=1.0\columnwidth]{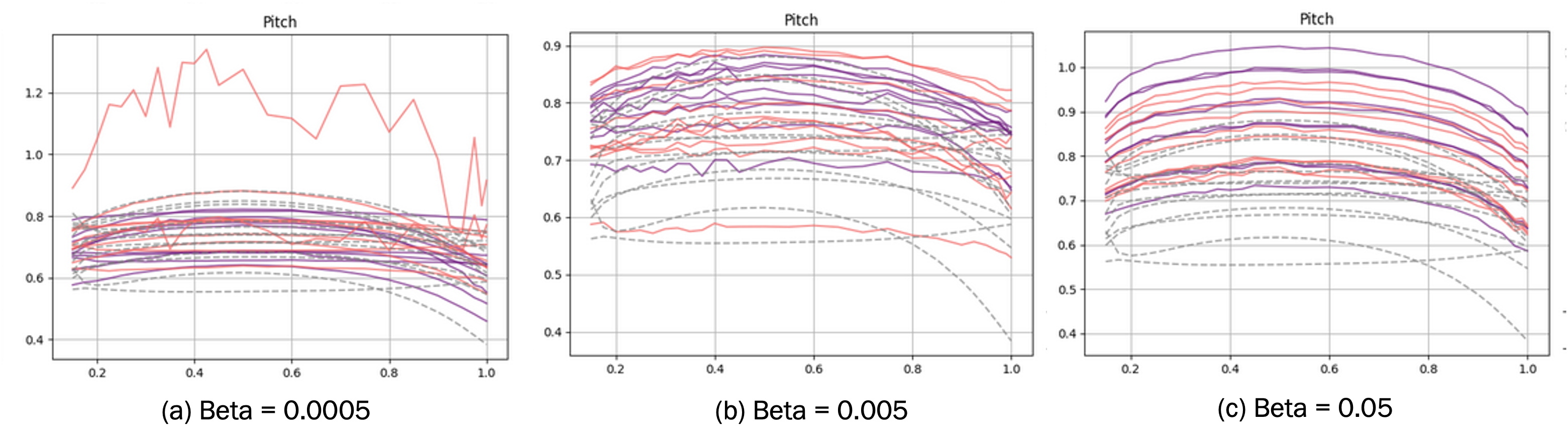}
\caption{Generated designs with different $\beta$ values in the VAE and $\epsilon$-prediction loss. $[J,K_T,\eta,D,B] = [0.6, 0.1, 0.75, 2.3, 4\ \text{or}\ 5]$ as condition. Blade number 4 is shown in purple; 5 in red.}
\label{result:fig:VAE_beta}
\end{figure}

\subsection{Latent Diffusion Model}

For the latent diffusion model, we adopt a two-stage approach: (1) a VAE is first trained to encode the 162-dimensional design vectors into a lower-dimensional latent space (64 dimensions); and (2) a conditional diffusion model is then trained in this latent space, conditioned on $\mathbf{c}_{\text{gen}} = [J, K_T, \eta, D, B]$, to generate new latent codes that are subsequently decoded back into propeller designs.

To study the effect of the latent regularization, we trained three VAE models with different $\beta_{\text{VAE}}$ values and compared reconstruction error and KL divergence. We then trained conditional diffusion models on the resulting latent spaces using two alternative loss formulations: the standard noise-prediction ($\epsilon$-prediction) loss and a velocity-prediction loss. Empirically, models trained with velocity prediction produced more accurate performance outputs and smoother, more physically plausible design geometries, whereas $\epsilon$-prediction models occasionally generated infeasible designs under edge-case conditions.

When using $\epsilon$-prediction, smaller $\beta_{\text{VAE}}$ values improved reconstruction accuracy of the VAE but made the latent space less regular, leading to unstable patterns in relatively sparse regions of the data distribution and, in some cases, to infeasible designs (Figure~\ref{result:fig:VAE_beta}). Larger $\beta_{\text{VAE}}$ values yielded a more regular latent space and consistently feasible designs, but at the cost of higher reconstruction error. These observations highlight the sensitivity of latent diffusion models to the quality and structure of the underlying latent representation, particularly when the diffusion network is trained to predict noise directly. Switching to velocity prediction largely mitigated these issues, stabilizing training and improving the robustness of generated designs across conditions.

Based on this exploration, the best-performing configuration consist of: (1) a VAE trained with $\beta_{\text{VAE}} = 0.0005$, which provided the lowest reconstruction error while retaining reasonable latent regularization; and (2) a conditional diffusion model trained in this latent space using the velocity-prediction loss.

\begin{figure}[!htb]
\centering
\includegraphics[width=0.45\columnwidth]{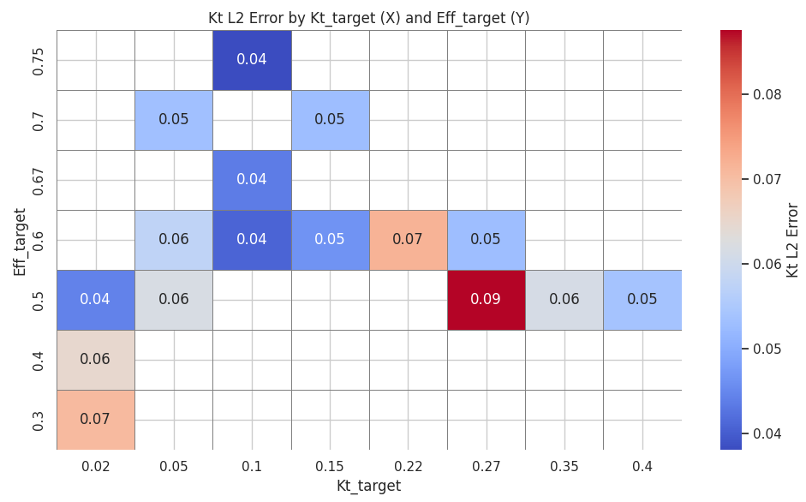}
\hfill
\includegraphics[width=0.45\columnwidth]{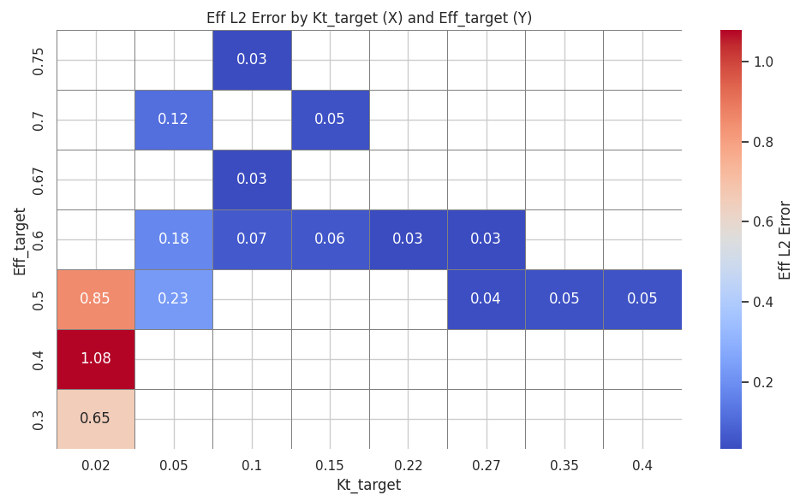}
\caption{(Left) relative $L_2$ error for $K_T$ across target conditions at $J=0.6$. (Right) relative $L_2$ error for $\eta$ across target conditions at $J=0.6$.}
\label{result:fig:LDM_errors}
\end{figure}

\begin{figure}[!htb]
\centering
\includegraphics[width=1.0\columnwidth]{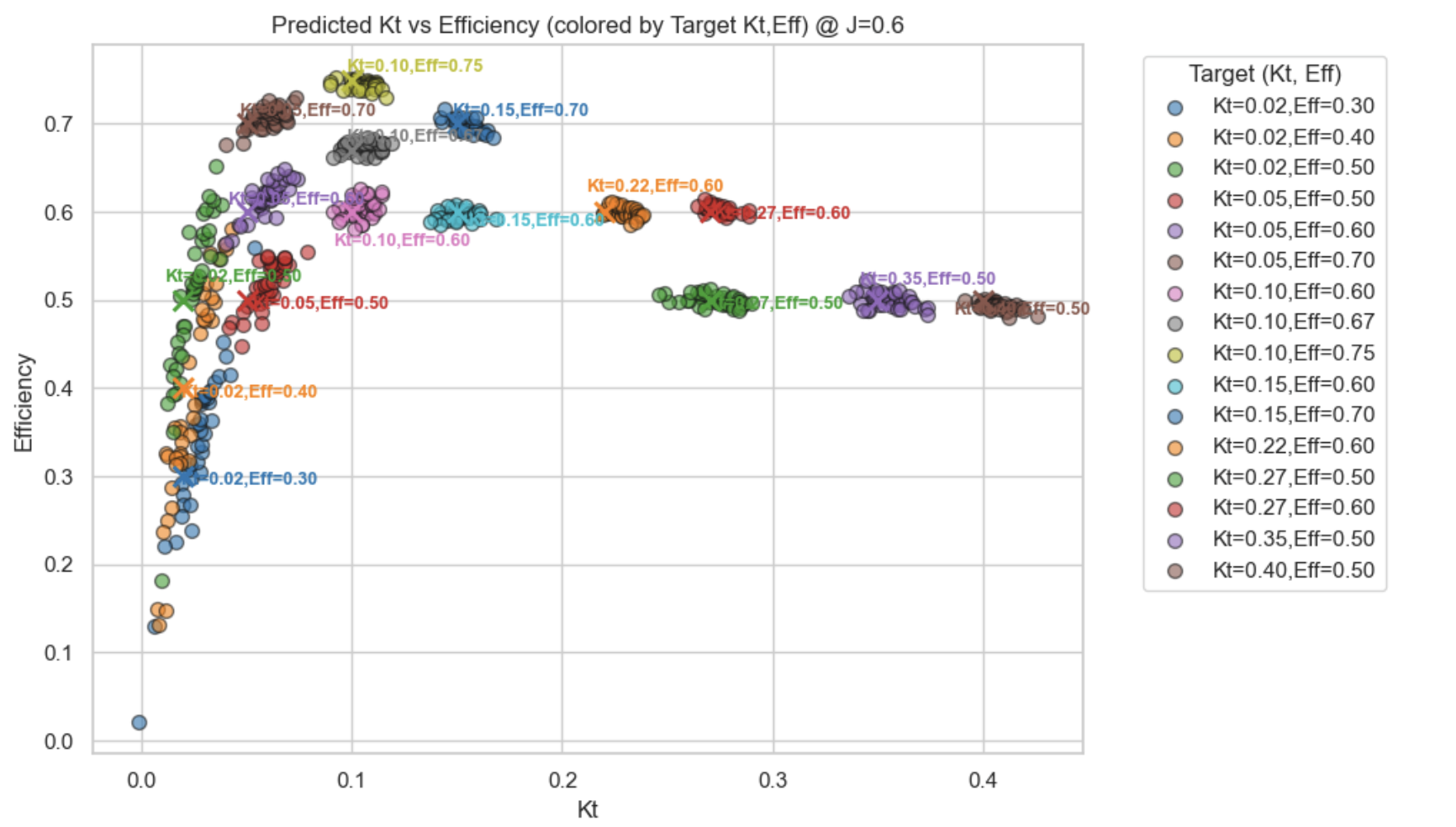}
\caption{Generated design sample distribution from the latent diffusion model.}
\label{result:fig:LDM_validation}
\end{figure}

The resulting performance across different conditioning targets is shown in Figure~\ref{result:fig:LDM_errors}. As with the cVAE, the largest discrepancies occur for highly loaded and low-efficiency conditions (higher $K_T$ and lower $\eta$), where the training data are sparse. This trend is further illustrated in Figure~\ref{result:fig:LDM_validation}, where deviations from the target become more pronounced in these regions of the operating space.

\subsubsection{Comparison with cVAE.}

To compare the two generative approaches, we evaluated designs produced by both the cVAE and the LDM using \textsc{PropElements}. Table~\ref{tab:generation_perf} summarizes the resulting relative $L_2$ errors in $K_T$ and $\eta$ with respect to the conditioning targets. Under the current configurations, the cVAE achieves slightly lower errors for both performance metrics, indicating marginally tighter control around the specified operating condition.

\begin{table}[!hb]
\centering
\caption{Performance comparison of cVAE and LDM (PropElement validation).}
\resizebox{\columnwidth}{!}{%
\begin{tabular}{lcccc}
\hline
Model & Mean $K_T$ Error \% & Mean $\eta$ Error \% & $K_T$ $L_2$ Error & $\eta$ $L_2$ Error \\
\hline
cVAE & 0.759 & 1.571 & 0.154 & 0.733 \\
LDM  & 1.266 & 2.671 & 0.243 & 1.137 \\
\hline
\end{tabular}%
}
\label{tab:generation_perf}
\end{table}

Beyond pointwise accuracy, we further quantify geometric diversity using two complementary metrics: spread coefficient (SC) and conditional novelty~\cite{khan2023shiphullgan}. Let $\{{x}_i^{(G)}\}_{i=1}^{m}$ denote the $m$ generated designs in the normalized design space $\mathbb{R}^d$, and let ${x}^{(G)}_{\text{centroid}} = \frac{1}{m}\sum_{i=1}^{m} {x}_i^{(G)}$ denote their centroid. The spread coefficient (SC) is defined as
\begin{equation}
    \mathrm{SC} = \frac{1}{m} \sum_{i=1}^{m} 
    \left\| {x}_i^{(G)} - {x}^{(G)}_{\text{centroid}} \right\|_2 ,
\end{equation}
which measures the average dispersion of generated samples around their centroid and reflects intra-set diversity. To assess how far generated samples deviate from the training manifold ${X} = \{{x}_j\}_{j=1}^{n}$, we compute the conditional novelty as
\begin{equation}
    \mathrm{Novelty} =
    \frac{1}{m} \sum_{i=1}^{m}
    \min_{1 \le j \le n}
    \left\| {x}_i^{(G)} - {x}_j \right\|_2 ,
\end{equation}
which represents the average nearest-neighbor distance from each generated design to the training dataset.

When generating $10$ samples per conditioning case, the cVAE produces $\mathrm{SC} \approx 6.69$ for both $4$- and $5$-blade configurations, with corresponding novelty values of $7.28$ and $8.72$, respectively, indicating that the samples remain close to the expert manifold (for reference, the expert designs exhibit a $\mathrm{SC}=6.59$ in the same normalized space). In contrast, the LDM exhibits substantially higher dispersion, achieving $\mathrm{SC} = 19.60$ and $23.76$ for the $4$- and $5$-blade cases, with novelty values of $14.36$ and $18.94$, respectively.

\begin{figure}[!htb]
\centering
\includegraphics[width=1\columnwidth]{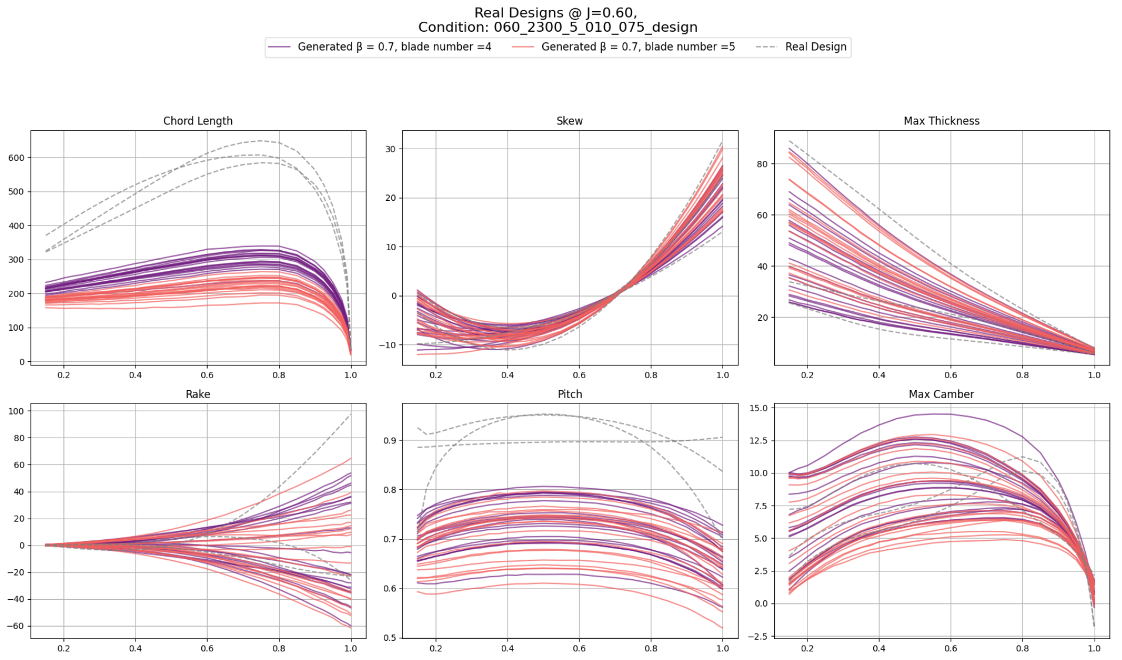}
\includegraphics[width=1\columnwidth]{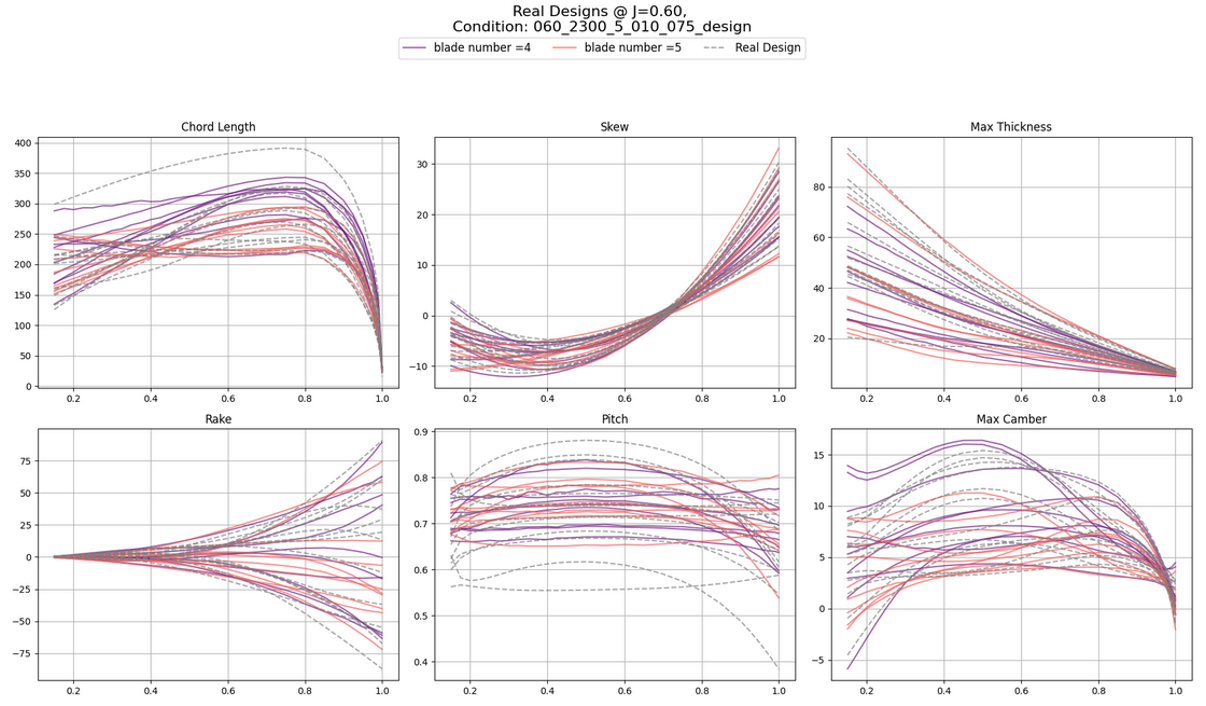}
\caption{Comparison of generated designs from (top) cVAE and (bottom) latent diffusion model at the same condition $[J,K_T,\eta,D,B] = [0.6, 0.1, 0.75, 2.3, 4\ \text{or}\ 5]$. Blade number 4 is shown in purple, 5 in red. Latent diffusion model produces more diverse designs than cVAE.}
\label{result:fig:high_Eff_samples}
\end{figure}

Notably, the LDM spread exceeds the intrinsic spread of the real dataset, indicating that it explores regions beyond the convex hull of expert designs. At the same time, hydrodynamic simulations confirm that the generated designs maintain performance close to the conditioning targets. This demonstrates that the LDM expands the geometric design space while remaining within the performance-feasible manifold. Figure~\ref{result:fig:high_Eff_samples} visually supports this quantitative observation. For chord length, rake, pitch, and maximum camber, the radial distributions of LDM-generated designs exhibit a wider spread and more dramatic curve compared to the cVAE, while preserving smooth and physically plausible profiles. 

These results suggest distinct generative behaviors. The cVAE primarily interpolates within the expert design manifold, producing compact and conservative variations. The LDM, by contrast, enables broader exploration and extrapolation in geometry space without compromising hydrodynamic feasibility. Consequently, the LDM is particularly advantageous for early-stage concept generation and design-space expansion, whereas the cVAE provides slightly tighter performance control around a specified operating point.

\begin{figure}[!htb]
\centering
\includegraphics[width=1.0\columnwidth]{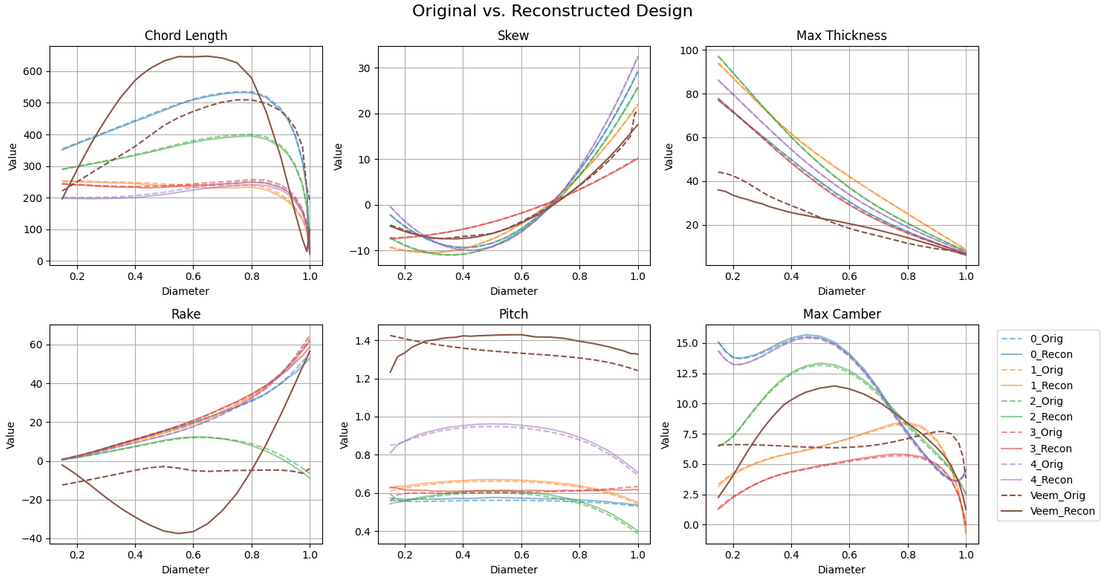}
\caption{Out-of-distribution design reconstruction. Solid lines represent original Veem designs (unseen by the model), dotted lines their VAE reconstructions. Numbers 1--5 indicate the nearest neighbors of the Veem design in latent space.}
\label{method:fig:ood_designs}
\end{figure}

\subsection{Limitation: Out-of-Distribution (OOD) Design}

To assess the robustness of the learned latent representation, we tested the VAE and latent diffusion model on out-of-distribution (OOD) geometries constructed from alternative expert design families (Figure~\ref{method:fig:ood_designs}). These OOD designs differ systematically from the training set in their chord, pitch, and skew distributions, and are intended to mimic propeller series or designer styles that were not present during data generation.

When encoded and reconstructed by the VAE, these OOD designs were consistently mapped back toward the manifold of in-distribution geometries rather than faithfully preserved. In other words, the autoencoder tended to “pull” unseen design families toward the dominant patterns observed in the training data, indicating limited generalization beyond the training distribution.

This behavior is expected given the finite scope of the dataset, but it underscores an important limitation of the present framework: generalization to genuinely new propeller families or radically different design philosophies is constrained by the coverage of the physics-based data generation pipeline. Addressing this limitation will require more diverse training datasets and potentially hybrid strategies, such as design refinement with evolutionary optimization, or incorporating reinforcement learning and active learning to deliberately expand the design space into underrepresented or novel regions.

\begin{table*}[!htb]
\centering
\caption{Summary of expert or reference propeller designs by vessel type.}
\label{tab:hull_summary}
\small
\setlength{\tabcolsep}{5pt}
\begin{tabularx}{\textwidth}{l c *{6}{>{\centering\arraybackslash}X}}
\toprule
Hull type & Count &
Hull length (m) & No. shafts &
Engine power (kW) & RPM &
Prop. diam. (mm) & No. blades \\
\midrule
Catamaran      & 4  & \SIrange{27}{60}{} & \SIrange{2}{4}{} & \SIrange{610}{1380}{} & \SIrange{607}{805}{}  & \SIrange{1100}{1295}{} & 5 \\
Crew Boat      & 10 & \SIrange{25}{42}{} & \SIrange{2}{3}{} & \SIrange{334}{1342}{} & \SIrange{649}{1033}{} & \SIrange{720}{1250}{}  & 5 \\
Landing Craft  & 4  & \SIrange{47}{75}{} & 2                & \SIrange{501}{759}{}  & \SIrange{284}{402}{}  & \SIrange{1600}{2000}{} & \SIrange{4}{5}{} \\
Tug Boat       & 4  & \SIrange{25}{30}{} & 2                & \SIrange{610}{759}{}  & \SIrange{234}{284}{}  & \SIrange{1800}{2250}{} & 4 \\
Utility Vessel & 8  & \SIrange{13}{49}{} & \SIrange{1}{3}{} & \SIrange{224}{2013}{} & \SIrange{430}{1040}{} & \SIrange{760}{1600}{}  & \SIrange{4}{5}{} \\
\bottomrule
\end{tabularx}
\end{table*}


\section{Conclusions}

We presented an AI-driven framework for performance-to-design generation of marine propellers, combining a physics-based dataset, a fast surrogate model, and conditional generative models (cVAE and latent diffusion). The surrogate accurately predicts $K_T$, $K_Q$, and $\eta$ in milliseconds, enabling real-time feedback and reducing reliance on repeated numerical solver calls. Both generative models produce realistic and hydrodynamically plausible propeller designs that broadly match target performance, with the latent diffusion model offering smoother feature trends and greater design diversity.

Two main limitations remain: limited ability to generate truly out-of-distribution designs and a feature-based geometry representation that is less intuitive than direct 3D meshes. In future work, we will apply the full end-to-end pipeline, from performance-to-design generation to evolutionary refinement, to practical crew-boat case studies and validate the resulting designs using CFD simulations


\section*{Acknowledgments}

This research is supported by A*STAR under its RIE2025 Industry Alignment Fund – Industry Collaboration Projects (IAF-ICP) I2201E0016 and Mencast Marine Pte Ltd, Singapore, and by the National Research Foundation, Singapore (NRF), Maritime and Port Authority of Singapore (MPA) and Singapore Maritime Institute (SMI) under its Maritime Transformation Programme (Project No. SMI-2024-MTP-01). Any opinions, findings and conclusions or recommendations expressed in this material are those of the author(s) and do not reflect the views of NRF, MPA and SMI.


\nocite{*}

\bibliographystyle{asmeconf}  
\bibliography{asmeconf-sample}

\appendix

\section{Expert and Reference Propeller Designs}
\label{app:reference_designs}

The expert and reference propeller designs used to derive the latent space (Section~3) comprise 30 propeller--vessel samples that were designed by marine propeller experts in response to client requests within the past 10 years and are already operational in service. The set spans five hull categories: Crew Boat (10), Utility Vessel (8), Catamaran (4), Tug Boat (4), and Landing Craft (4). They are summarized in Table~\ref{tab:hull_summary}. These expert designs provide physically plausible anchors for latent-space learning within the targeted operating regime.

\section{Blade Thickness Constraint}
\label{app:thickness_constraint}

In the evolutionary refinement stage, blade thickness is enforced as a \emph{hard} constraint following the classification-rule minimum thickness propeller strength requirements~\cite{veritas2015rules}. The blade thickness of a propeller at the reference radii must satisfy
\begin{equation}
    \small
    t_{0.25} \;\ge\; 3.2\, \left[f 
    \frac{1.5\times 10^{6}\,\rho_{0.25}\, M_T \;+\; 51\,\delta \left(\frac{D}{100}\right)^{3} BAR\, l_{0.25}\, RPM^{2}\, h}
    {l_{0.25}\, B\, R_m}
    \right]^{1/2},
    \label{eq:bv_t025}
\end{equation}
\begin{equation}
    \small
    t_{0.6} \;\ge\; 1.9\, \left[f 
    \frac{1.5\times 10^{6}\,\rho_{0.6}\, M_T \;+\; 18.4\,\delta \left(\frac{D}{100}\right)^{3} BAR\, l_{0.6}\, RPM^{2}\, h}
    {l_{0.6}\, B\, R_m}
    \right]^{1/2},
    \label{eq:bv_t06}
\end{equation}
where \(t_{0.25}\) and \(t_{0.6}\) are the blade thicknesses (mm) at radii \(0.25R\) and \(0.6R\), respectively. The remaining symbols: \(f\) is a material factor, \(R_m\) is the minimum tensile strength (N/mm\(^2\)),
\(\delta\) is the material density (kg/dm\(^3\)), \(B\) is the number of blades, \(D\) is the propeller diameter (m), \(BAR\) is the developed area ratio,
\(RPM\) is the rotational speed (rev/min), \(h\) is the rake (mm), \(l_{0.25}\) and \(l_{0.6}\) are the expanded widths (mm) at \(0.25R\) and \(0.6R\).
The pitch ratios are \(\rho_{0.25} = D/H\) and \(\rho_{0.6}=D/H_{0.6}\), with \(H\) the mean pitch (m) and \(H_{0.6}\) the pitch at \(0.6R\) (m).
The transmitted torque \(M_T\) (kN$\cdot$m) may be taken as
\begin{equation}
    M_T = 9.55\left(\frac{P}{RPM}\right),
    \label{eq:bv_torque}
\end{equation}
with \(P\) the maximum continuous propulsion power (kW). Material parameters \((R_m, \delta, f)\) are selected according to the material~\cite{veritas2015rules}, including common bronze (400, 8.3, 7.6), manganese bronze (440, 8.3, 7.6), nickel-manganese bronze (440, 8.3, 7.9), aluminum bronze (590, 7.6, 8.3), and steel (440, 7.9, 9.0).





\end{document}